\documentclass[preprint,amsmath,showpacs,amssymb,aps,nofootinbib,floatfix]{revtex4-1}
\usepackage{epsfig,amsmath,amssymb}
\usepackage{dcolumn,color}
\usepackage{bm}
\usepackage{graphicx}

\begin{document}
\bibliographystyle{unsrt}

\title{Differential flow correlations in relativistic heavy-ion collisions}
\author{Jing Qian$^{1, *}$, Ulrich Heinz$^2$, Ronghua He$^1$ and Lei Huo$^1$}
\affiliation{$^1$ Department of Physics, Harbin Institute of Technology, Harbin, 150001, People's Republic of China}
\affiliation{$^2$ Department of Physics, The Ohio State University, Columbus, OH 43210-1117, USA}
\email[Correspond to\ ]{qianjing8758@gmail.com}

\begin{abstract}
A systematic analysis of correlations between different orders of $p_T$-differential flow is presented, including mode coupling effects in flow vectors, correlations between flow angles (a.k.a. event-plane correlations), and correlations between flow magnitudes, all of which were previously studied with integrated flows. We find that the mode coupling effects among differential flows largely mirror those among the corresponding integrated flows, except at small transverse momenta where mode coupling contributions are small. For the fourth- and fifth-order flow vectors $V_4$ and $V_5$ we argue that the event plane correlations can be understood as the ratio between the mode coupling contributions to these flows and and the flow magnitudes. We also find that for $V_4$ and $V_5$ the linear response contribution scales linearly with the corresponding cumulant-defined eccentricities but not with the standard eccentricities. 

\end{abstract}

\date{\today}

\maketitle

\section{Introduction}
\label{sec1}

The ultimate goal of studying relativistic heavy ion collisions is to extract from experimentally measured final particle momentum distributions quantitatively precise information on the transport properties and dynamical evolution of the quark-gluon plasma (QGP) generated in these collisions. The azimuthal anisotropy of particle emission in the transverse plane, known as anisotropic flow, is one key observable suggesting that QGP behaves like an almost perfect liquid \cite{Heinz:2013th}. Using an azimuthal Fourier expansion of the single particle distribution up to harmonic order $N$, this anisotropy can be characterized by $2N$ parameters: the flow magnitude $v_n$ and the flow angle $\Psi_n$ relative to the reaction plane which is often called $n$-th order event plane angle ($1\leq n\leq N$). They are combined into the complex flow vectors $V_n=v_n e^{in\Psi_n}$.

Due to quantum mechanical fluctuations in the initial conditions created in heavy-ion collisions, the flow vectors $V_n$ fluctuate from event to event, even for identical impact parameters and collision systems. Correlations between anisotropic flow vectors of different orders have been studied both theoretically and experimentally. Examples are correlations between flow angles (a.k.a, event-plane correlations) \cite{Teaney:2010vd, Teaney:2012gu, Jia:2012ma, Jia:2012ju, Aad:2014fla, Qiu:2012uy}, correlations between the magnitudes of the flow harmonics \cite{Aad:2015lwa, Qian:2016pau, ALICE:2016kpq}, and nonlinear mode coupling effects between flow vector contributing to $V_n$ for $n>3$ \cite{Yan:2015jma, Qian:2016fpi}. These correlations may shed light on the fluctuating initial conditions, but their strength is also affected by dissipative effects ion the dynamical evolution of the QGP.\footnote{%
 		Even without E-b-E fluctuations, the even orders of flow are correlated because of the almond 
		shaped deformation of the initial spatial distribution in noncentral collisions which is not of pure 
		$\cos(2\phi)$ form.}

Since measurements of such correlations are very statistics-hungry, existing correlation studies are almost exclusively for the integrated flows, using the flow vectors of all charged particles in a centarin rapidity window without differentiating them according to their transverse momentum. Given the continuously increasing number of collected data we here ask the question what would change if one performed this studies differentially in transverse momentum.  We present a systematic study of the correlation between differential flows of differerent harmonic orders for Pb+Pb collisions at 2.76 TeV, using the VISH2+1 hydrodynamic code \cite{Song:2007fn, Shen:2014vra} to describe the dynamical evolution of the collision. Admittedly, our study does not suffer from the same kind of statistical limitations faced by experimentalists: We use the Cooper-Frye prescription to compute the final particle spectra from the hydrodynamic output, which yields a continuous final particle momentum distribution, corresponding to the limit of an infinite number of particles emitted from each event. Therefore, our results are affected only by fluctuations associated with the fluctuating initial conditions (which we sample by evolving 2000 events per centrality class through the hydrodynamic code) and not by finite number statistical fluctuations in the final state that arise in real experiments from the fact that Nature, due to the limited energy content of each event, can sample the final momentum distribution only with a finite number of particles. -- We here use the MC-Glauber initial conditions as input, start the hydrodynamic evolution at $\tau_0=0.6 $ fm/$c$ without pre-equilibrium flow, and end it on an isothermal freeze-out surface of temperature $T_\mathrm{dec}=120$ MeV. Unless otherwise stated, the shear viscosity is set to a default value of $\eta/s=0.08$. Correlations between flow angles and flow magnitudes are discussed in Secs.~\ref{sec2} and \ref{sec3}, respectively. In Sec.~\ref{sec4} we present the mode coupling effects in higher order of differential flow vectors. The results are discussed and summarised in Sec.~\ref{sec5}. Some studies elucidating the meaning of the linear part (previously called the ``linear response part'') of higher harmonic flows are presented in the Appendix.

\section{Correlations between flow angles}
\label{sec2}

Correlations between different flow angles are usually called event plane correlations whereas the correlations between the angles associated with the corresponding initial spatial eccentricities are known as participant plane correlations. These multi-plane correlations have yielded insight into the initial conditions and hydrodynamic evolution of heavy ion collisions \cite{Teaney:2010vd, Teaney:2012gu, Jia:2012ma, Jia:2012ju, Aad:2014fla, Qiu:2012uy}.

To calculate the event plane correlations we use the scalar product definition which does not depend on the event-plane resolution \cite{Bhalerao:2013ina}: 
\begin{equation}
  \langle \mathrm{cos} (c_1 n_1 \phi_{n_1} + ... + c_k n_k \phi_{n_k})\rangle\{\mathrm{SP}\} 
  := \frac{\langle Q_{n_1}^{c_1} Q_{n_2}^{c_2} ... Q_{n_k}^{c_k}\rangle}
             {\sqrt{\langle Q_{n_1}^{c_1} Q_{n_1}^{*c_1} \rangle ... \langle Q_{n_k}^{c_k} Q_{n_k}^{*c_k} \rangle}}.
\label{EP_SP}
\end{equation}
Here, $\langle \mathrm{cos}(4(\Psi_2-\Psi_4)) \rangle\{\mathrm{SP}\} $ is chosen as an example of a two-plane correlation, and $\langle \mathrm{cos}(2\Psi_2 + 3\Psi_3 - 5\Psi_5)) \rangle\{\mathrm{SP}\} $ as an example of a three-plane correlation:
\begin{equation}
\begin{split}
	\langle \mathrm{cos}(4(\Psi_2-\Psi_4)) \rangle\{\mathrm{SP}\}  
	&:= \frac{\langle V_4 V_2^{*2} \rangle}{\sqrt{\langle v_4^2 \rangle \langle v_2^4 \rangle}},
\\
	\langle \mathrm{cos}(2\Psi_2 + 3\Psi_3 - 5\Psi_5)) \rangle\{\mathrm{SP}\}  
	&:= \frac{\langle V_5 V_2^* V_3^* \rangle}
	              {\sqrt{\langle v_5^2 \rangle \langle v_2^2 \rangle \langle v_3^2 \rangle}}.
\end{split}
\label{EP45}
\end{equation}
From this definition we see that $\langle \mathrm{cos}(4(\Psi_2{-}\Psi_4)) \rangle\{\mathrm{SP}\}$ in fact equals the Pearson correlation coefficient between $V_4$ and $V_2^2$. On the other hand, $\langle \mathrm{cos}(2\Psi_2{+}3\Psi_3{-}5\Psi_5)) \rangle\{\mathrm{SP}\}$ does not equal the Pearson correlation coefficient between $V_5$ and $V_2V_3$ unless the fluctuations of the elliptic and triangular flows are uncorrelated and factorize as follows: $\langle v_2^2 v_3^2 \rangle=\langle v_2^2 \rangle \langle v_3^2 \rangle$. While this may be a reasonable assumption for $v_2$ and $v_3$ because $v_3$ is dominated by initial state fluctuations whereas $v_2$ has generally a strong geometric component, it is certainly not justifiable for the correlation between $v_2^2$ and $v_4^2$ which are correlated with each other by the deformed initial collision geometry in non-central collisions. This geometric correlation between the elliptic and quadrangular flow magnitudes affects the three-plane correlation $\langle \mathrm{cos}(2\Psi_2 + 4\Psi_4 - 6\Psi_6)) \rangle\{\mathrm{SP}\}$. This shows that event-plane correlations in general are not Pearson correlation coefficients between the corresponding $V_n$.

%
\begin{figure}[!hbt]
    \includegraphics[width=\linewidth]{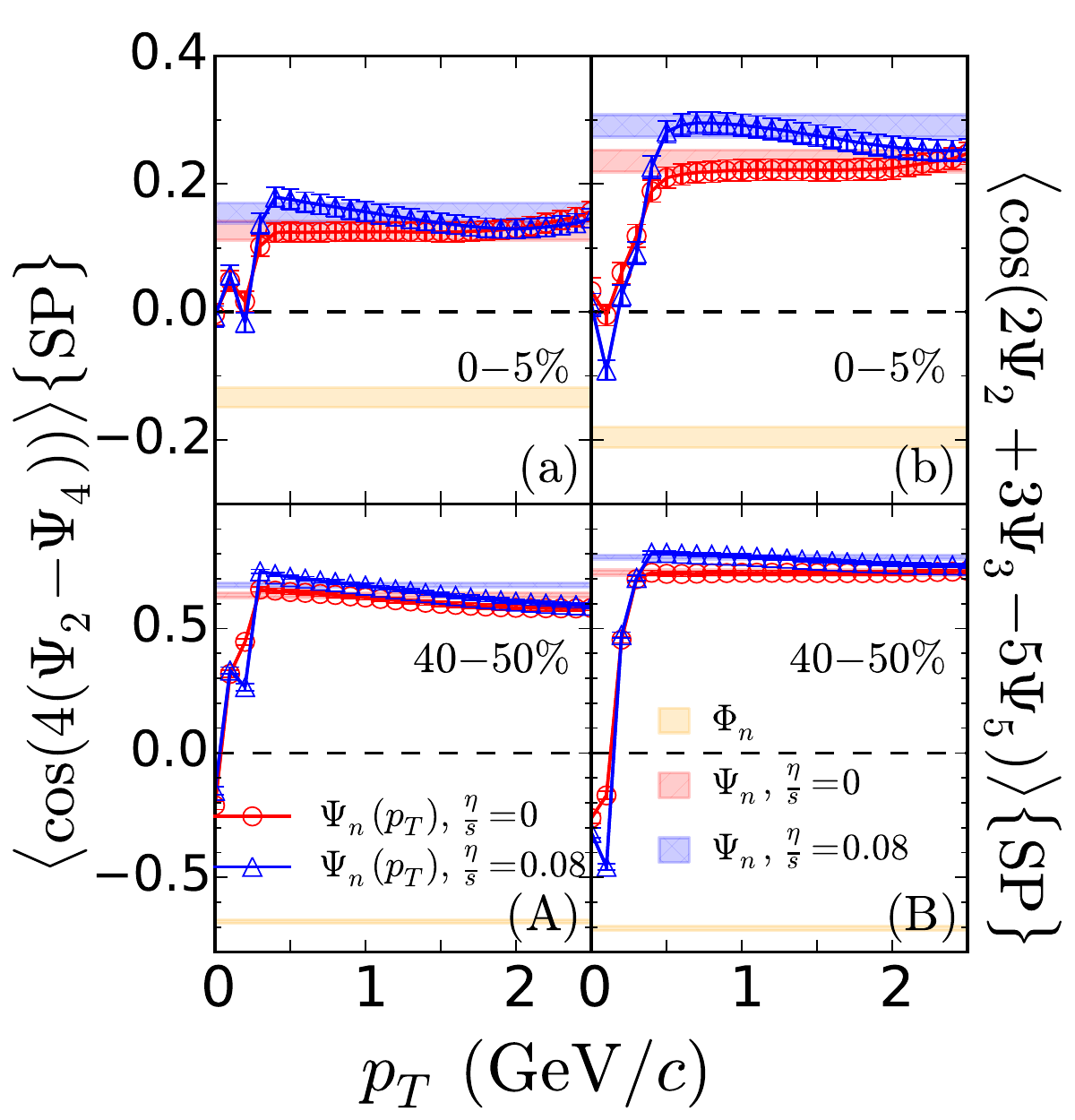}
    \caption{(Color online)
    \baselineskip=14pt
    $p_T$ dependence of the two- and three-plane correlations 
    $\langle \mathrm{cos}(4(\Psi_2{-}\Psi_4)) \rangle\{\mathrm{SP}\}$ and three-plane correlation 
    $\langle \mathrm{cos}(2\Psi_2{+}3\Psi_3{-}5\Psi_5) \rangle\{\mathrm{SP}\}$.
    Subgraphs (a-b) are for events of 0-5\% centrality while (A-B) are for 40-50\% centrality. 
    In each centrality class 2000 hydrodynamic events are used for the analysis.
    Red and blue represent ideal and viscous hydrodynamic results, respectively.
    Markers show event-plane correlations for differential flows, and colored bands are those for 
    the corresponding integrated flows shown for comparison.     
    The corresponding participant plane correlations are shown as yellow bands. 
    The widths of the bands indicate the corresponding variances due to event-by-event fluctuations.
    \label{F-II1}
}
\end{figure}
%

The event plane correlations (\ref{EP45}) for differential (as a function of transverse momemtum) and integrated flows are shown in Fig.~\ref{F-II1} and compared with the corresponding participant plane correlations. Note that for the $p_T$-differential event plane correlators all correlated particles are taken from the same $p_T$ bin. (This is different from the usual definition of second- or third-order $p_T$-differential flow cumulants (see e.g. \cite{Betz:2016ayq}) where only one particle comes from the selected $p_T$ bin and its flow vector is correlated with others constructed from all charged hadrons.) Red circles denote event-plane correlations of differential flows from ideal hydrodynamic evolution, and the red shaded bands show the corresponding $p_T$-integrated event-plane correlations of integrated for the same events for comparison. Blue triangles and blue shaded bands show the analogous correlations for events that were evolved with viscous hydrodynamics using shear viscosity $\eta/s=0.08$. The upper (lower) panels are for
central (0-5\% centrality) and mid-peripheral (40-50\% centrality) collisions. 

Similar to what was observed for the $p_T$-integrated flows, the event plane correlations between the $p_T$-differential flows increase with impact parameter from central to semi-peripheral collisions. Shear viscosity increases the strength of the event-plane correlation as previously found by Yan \cite{Teaney:2012gu} and Qiu \cite{Qiu:2012uy} for $p_T$- integrated flows. The shear viscous strengthening of the correlations is relatively more pronounced in central than in more peripheral collisions, but it appears to disappear at higher $p_T$ values.  At low $p_T$, the differential event plane correlations are much weaker than those of the integrated flows and almost negligible. Increasing with $p_T$, the differential event plane correlations become approximately equal to those of the integrated flows around $p_T \approx 0.5$ GeV/$c$; at even larger $p_T$, however, for viscous evolution the $p_T$-differential correlations drop below the $p_T$-integrated values, indicating that the viscous strengthening of the event-plane correlations operates only at thermal transverse momenta and disappears for harder $p_T$. 

The differences between differential and integrated event plane correlations can to a large extent be explained as a consequence of the decorrelation between the flow angles at different $p_T$-values. As noted in \cite{Heinz:2013bua}, 
in general the flow angle $\Psi_n$ depend on $p_T$ and, as a function of $p_T$, $\Psi_n(p_T)$ wanders around the `average angle' $\Psi_n$. The $p_T$-averaged event-plane correlator thus closely represents the $p_T$-differential one only in the $p_T$-region in which the majority of particles are emitted. At small $p_T$, the variance $\sigma(\Psi_n(p_T) - \Psi_n)$ is quite large for all $n$ due to the fluctuating $\Psi_n(p_T)$. This is the likely reason for  the much weaker event-plane correlations of the $p_T$-differential flows at small $p_T$ compared to the $p_T$-averaged ones: At small $p_T$, the directions of the complex flow vectors are almost uncorrelated \cite{DiFrancesco:2016srj}. For this reason, we will mostly ignore the low-$p_T$ region in the rest of the paper.

\section{Correlations between flow magnitudes}
\label{sec3}

There are mainly two ways to describe the correlations between flow magnitudes: One way is to study the correlation between $v_m$ and $v_n$ via the event-shape selection method, suggested by the ATLAS collaboration \cite{Jia:2014jca, Aad:2015lwa} and already tested with the hydrodynamic model \cite{Qian:2016pau}. The other way is using the Symmetric 2-harmonic 4-particle Cumulant (or Moment) $SC(m, n)$ to evaluate the correlation between $v_m$ and $v_n$, suggested by the ALICE Collaboration \cite{ALICE:2016kpq} and tested with hydrodynamic, transport and hybrid models. In particular, the authors of \cite{Zhu:2016puf} studied the $p_T$ dependence of the normalized correlators $SC(m, n)$, ie. $NSC^v(m,n)\equiv\frac{\langle v_m^2 v_n^2 \rangle-\langle v_m^2 \rangle \langle v_n^2 \rangle}{\langle v_m^2 \rangle \langle v_n^2 \rangle}$, between the magnitudes of the differential flows $v_m(p_T)$ and $v_n(p_T)$ at 20-30\% centrality with the VISH2+1 and AMPT models. They found that for both models, $NSC^v(3, 2)$ and $NSC^v(4, 3)$ change sign from negative to positive with increasing $p_T$ around at $p_T{\,\sim\,}3$\,GeV/$c$.

Furthermore, Niemi and collaborators studied the linear correlation coefficient $c(v_m, v_n)$ of the differential flows  $v_m(p_T)$ and $v_n(p_T)$ as a function of transverse momentum for 20-30\% Au+Au collisions at $\sqrt{s_\mathrm{NN}}=200$ GeV with the hydrodynamic model \cite{Niemi:2012aj}. In their calculation, $c(v_2, v_3)$ and $c(v_3, v_4)$ also exhibit a sign change with increasing $p_T$. However, they argued that at 20-30\% centrality, $(v_2, v_3)$ and $(v_3, v_4)$ are not linearly correlated since $c(v_2, v_3) \approx c(v_3, v_4) \approx 0$ . They also suggested that the differential $c(v_2, v_4)(p_T)$ is sensitive to shear viscosity and decoupling temperature and is strongly affected by $c(\epsilon_2, \epsilon_4)$ in the initial state.

In this paper, we use the event-shape selection method to study the correlations between differential flows \cite{Jia:2014jca}. Building on previous work reported in \cite{Qian:2016pau}, 42000 hydrodynamically generated events were divided into 14 equal centrality classes according to multiplicity, then ordered by $q_n$ and subdivided by percentile into 6 bins per centrality class ($0{-}0.1,\, 0.1{-}0.2,\, 0.2{-}0.5,\, 0.5{-}0.8,\, 0.8{-}0.9,$ and $0.9{-}1.0$) where $\bm{q}_n= q_n\,e^{in\Psi^q_n} \equiv \langle m_T e^{in\phi_p}\rangle/\langle m_T\rangle$ (with $m_T = \sqrt{m^2{+}p_T^2}$). The differential flow magnitude is calculated as $v_n\{2\}(p_T)=\langle v_n(p_T) \,v_n \mathrm{cos}\bigl(n(\Psi_n(p_T)-\Psi_n)\bigr) \rangle /\sqrt{\langle v_n^2 \rangle}$; here $v_n$ is $n$-th order integrated flow coefficient and $\langle ... \rangle$ denotes the average over events in one $q_2$ (or $q_3$) bin.

Before discussing the correlations between the differential flow magnitudes further we would like to emphasize that the event shape selection based on $q_n$ yields equivalent event classes for different $p_T$ ranges. As seen in Fig.~6 in \cite{Aad:2015lwa} for $n=2$ and 3, $v_n$ for the range $0.5 < p_T < 2$ GeV shows approximate linearity with $v_n$ in the range $3 < p_T < 4$ GeV, for different $q_n$ bins and in all centrality classes. 

%
\begin{figure*}[!hbt]
    \includegraphics[width=\linewidth]{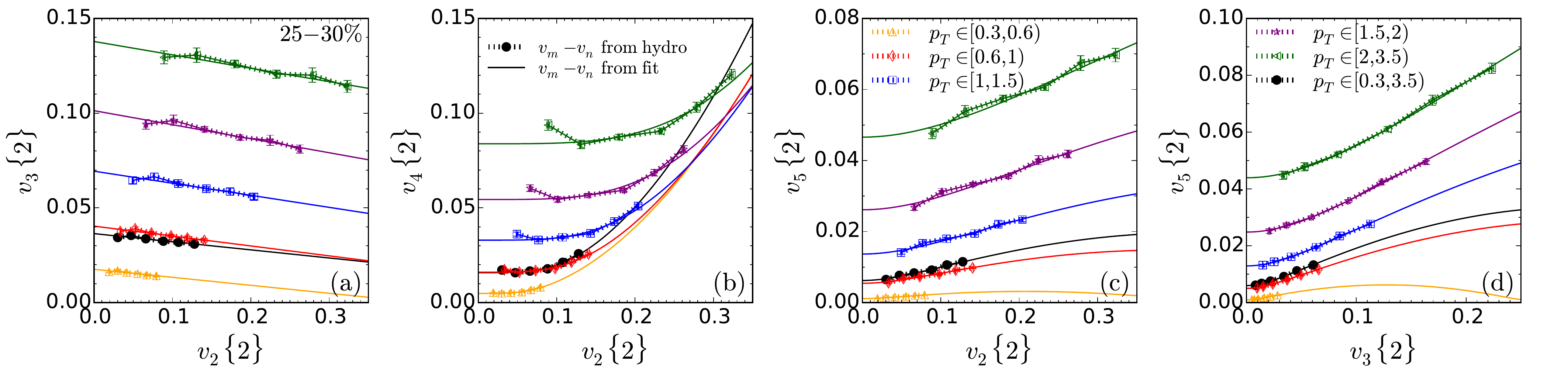}
    \caption{(Color online)
    \baselineskip=14pt
    $v_m\{2\}$ vs. $v_n\{2\}$ for different $p_T$ ranges at 25-30\% centrality, calculated with viscous 
    hydrodynamics using $\eta/s=0.08$. 3000 events in each 5\% centrality bin are used for the analysis.
    While solid black circles represent the integrated $v_m\{2\}-v_n\{2\}$ correlations, the colored hollow 
    markers show the correlations at different $p_T$. Markers connected by dotted lines represent the
    hydrodynamic results; the solid lines are fits using Eqs.~(\ref{vmvn}).
     \label{F-III1}
}
\end{figure*}
%

Fig.~\ref{F-III1} shows the correlations between $v_m$ and $v_n$ for different $p_T$ ranges, at 25-30\% collision centrality. Differently colored lines represent flows calculated within different $p_T$ ranges while different points along a line of given color represent different $q_n$ bins ($q_2$ bins for the left three panels, $q_3$ bins for the right panel). The solid black circles connected by dotted lines are the $p_T$-integrated $v_m-v_n$ correlations for comparison.

Since the differential $v_m$ and $v_n$ all increase with $p_T$ in the $p_T$ ranges shown here, and the integrated flows are the yield-weighted averages of the differential flows, the $p_T$-integrated black lines are in the middle of the colored lines representing $p_T$-differential correlations. In fact, the $p_T$-integrated $v_m-v_n$ correlation is quite close to the differential one for the range $0.6 < p_T < 1$ GeV/$c$. As for the $p_T$-integrated flows \cite{Aad:2015lwa, Qian:2016pau}, the differential $v_4$ and $v_5$ are positively correlated with the differential $v_2$ whereas $v_3$ is anticorrelated with $v_2$. In Ref.~\cite{Qian:2016pau} the following fit functions were found to yield good representations of the $p_T$-integrated $v_m-v_n$ correlations:
\begin{equation}
\label{vmvn}
\begin{split}
  v_3\{2\}&=v_3^0+k_3 v_2\{2\},\\
 v_4\{2\}&=\sqrt{(v_4^0)^2+(k_4 v_2^2\{2\})^2},\\
 v_5\{2\}&=\sqrt{(v_5^0)^2+(k_5 v_2\{2\} v_3\{2\})^2}.\\
\end{split}
\end{equation}
The solid colored lines in Fig.~\ref{F-III1} show that these fit functions represent the $p_T$-differential correlations equally well. 
%
\begin{figure}[!hbt]
    \includegraphics[width=\linewidth]{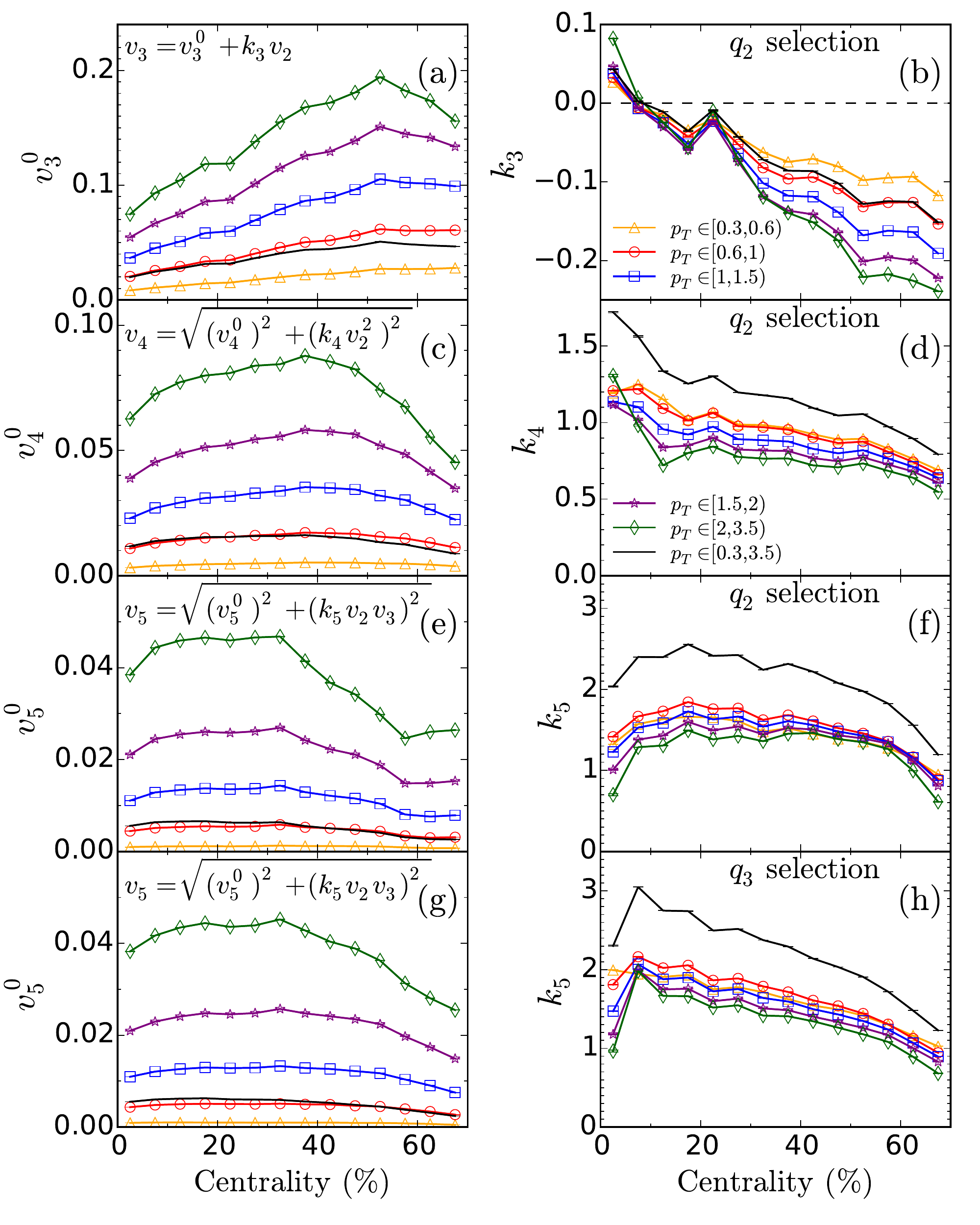}
    \caption{(Color online)
    \baselineskip=14pt
    Centrality dependence of the parameters $v_n^0$ and $k_n$ extracted from the fits of the 
    $v_m\{2\}-v_n\{2\}$ correlations shown in Fig.~\ref{F-III1} with Eqs.~(\ref{vmvn}), for the different 
    $p_T$ ranges (colored markers)) as well as the $p_T$-averaged correlation (black lines) shown in 
    that figure, using the same color coding.
     \label{F-III2}
}
\end{figure}
%
The corresponding fit parameters are plotted as functions of centrality in Fig.~\ref{F-III2}, $v_n^0$ in the left and $k_n$ in the right panels. One sees that the fit parameters for the differential flow correlations at different $p_T$ have similar centrality dependences as those for the integrated flows. Except for the most central collisions, $k_n$ decreases with impact parameter. $k_3$ is negative, due to the anti-correlation between $v_3$ and $v_2$ except for the most  central collisions. As discussed in \cite{Qian:2016pau} the latter is caused by neglecting $p{-}p$ multiplicity fluctuations in the initial conditions used in this study. $k_4$ and $k_5$ are both positive. Using $q_2$ or $q_3$ in the event-shape selection leads to some differences in the fitted parameters for $v_5$. The new information from the $p_T$-differential analysis is that $v_n^0$ increases while $k_n$ decreases with increasing $p_T$, for all $n$. This means that, as $p_T$ increases, the linear contribution $v_n^0(p_T)$ increases in sync with $v_n(p_T)$ whereas the strength of the non-linear mode coupling described by $k_n(p_T)$ decreases. Another interesting observation is that $k_4$ and $k_5$ of the $p_T$-integrated flows are larger than those of the $p_T$-differential flows, in all $p_T$ ranges and at all collision centralities. We will discuss this further in the next section.

\section{Mode coupling effects in the differential flow vectors}
\label{sec4} 

It has now been established that, while $V_2$ and $V_3$ respond almost linearly to their corresponding initial eccentricity vectors, $V_4$ and higher harmonic flows are affected by significant nonlinearities in their response. In \cite{Yan:2015jma, Qian:2016fpi}, $V_n\ (n>3)$ was decomposed into linear response and nonlinear mode coupling contributions as follows:
\begin{equation}
\label{VnDecomposition}
\begin{split}
  V_4 & = V_{4 L} + \chi_{422} V_2^2,\\
  V_5 & = V_{5 L} + \chi_{523} V_2 V_3,\\
  V_6 & = V_{6 L} + \chi_{624} V_2 V_{4L} + \chi_{633} V_3^2 + \chi_{6222}  V_2^3,\\
  V_7 & = V_{7 L} + \chi_{725} V_2 V_{5L} + \chi_{734} V_3 V_{4L} + \chi_{7223}  V_2^2 V_3.
\end{split}
\end{equation}
Some questions about the interpretation of $V_{nL}$ as the linear response contribution to the corresponding initial eccentricity were raised in Ref.~\cite{Qian:2016fpi}. In the Appendix we contribute to the further clarification of this question by showing empirically that for $n=4$ and 5 $V_{nL}$ responds approximately linearly to the cumulant-based but not to the moment-based initial eccentricities, as first suggested in \cite{Teaney:2012ke,Teaney:2013dta}. 
The mode coupling coefficients in Eqs.~(\ref{VnDecomposition}) are defined by
\begin{eqnarray}
\label{chi}
&&\chi_{422} = \frac{\mathrm{Re} \langle V_4 (V_2^*)^2 \rangle}{\langle v_2^4 \rangle}, \quad
\chi_{523} = \frac{\mathrm{Re} \langle V_5 V_2^* V_3^* \rangle}{\langle v_2^2 v_3^2 \rangle},
\nonumber\\
  &&\chi_{624} =
        \mathrm{Re}\frac{\langle V_6V_2^*V_4^* \rangle \langle v_2^4 \rangle 
                                  - \langle V_6 V_2^{*3} \rangle \langle V_4 V_2^{*2} \rangle}
                                   {\bigl( \langle v_4^2 \rangle \langle v_2^4 \rangle%
                                  {-}\langle V_4 V_2^{*2} \rangle^2\bigr)\,\langle v_2^2 \rangle},
\nonumber\\
  &&\chi_{633} = \frac{\mathrm{Re} \langle V_6 V_3^{*2} \rangle}{\langle v_3^4 \rangle},\quad
\chi_{6222} = \frac{\mathrm{Re} \langle V_6 V_2^{*3} \rangle}{\langle v_2^6 \rangle},
\\
  &&\chi_{725} = 
        \mathrm{Re}\frac{\langle V_7 V_2^* V_5^* \rangle \langle v_2^2 v_3^2 \rangle 
                                   - \langle V_7 V_2^{*2} V_3^*\rangle \langle V_5 V_2^* V_3^* \rangle}
                                   {\bigl(\langle v_5^2 \rangle \langle v_2^2 v_3^2 \rangle%
                                     {-}\langle V_5 V_2^* V_3^* \rangle^2\bigr)\,\langle v_2^2 \rangle},
\nonumber\\
  &&\chi_{734} = 
        \mathrm{Re}\frac{\langle V_7 V_3^* V_4^* \rangle \langle v_2^4 \rangle 
                                   - \langle V_7 V_2^{*2} V_3^* \rangle \langle V_4 V_2^{*2} \rangle}
                                   {\bigl(\langle v_4^2 \rangle \langle v_2^4 \rangle%
                                   {-}\langle V_4 V_2^{*2}\rangle ^2\bigr)\,\langle v_3^2 \rangle},
\nonumber\\
  &&\chi_{7223} = \frac{\mathrm{Re} \langle V_7 V_2^{*2} V_3^* \rangle}{\langle v_2^4 v_3^2 \rangle}.\nonumber
\end{eqnarray}
%
%
\begin{figure*}[!hbt]
    \includegraphics[width=\linewidth]{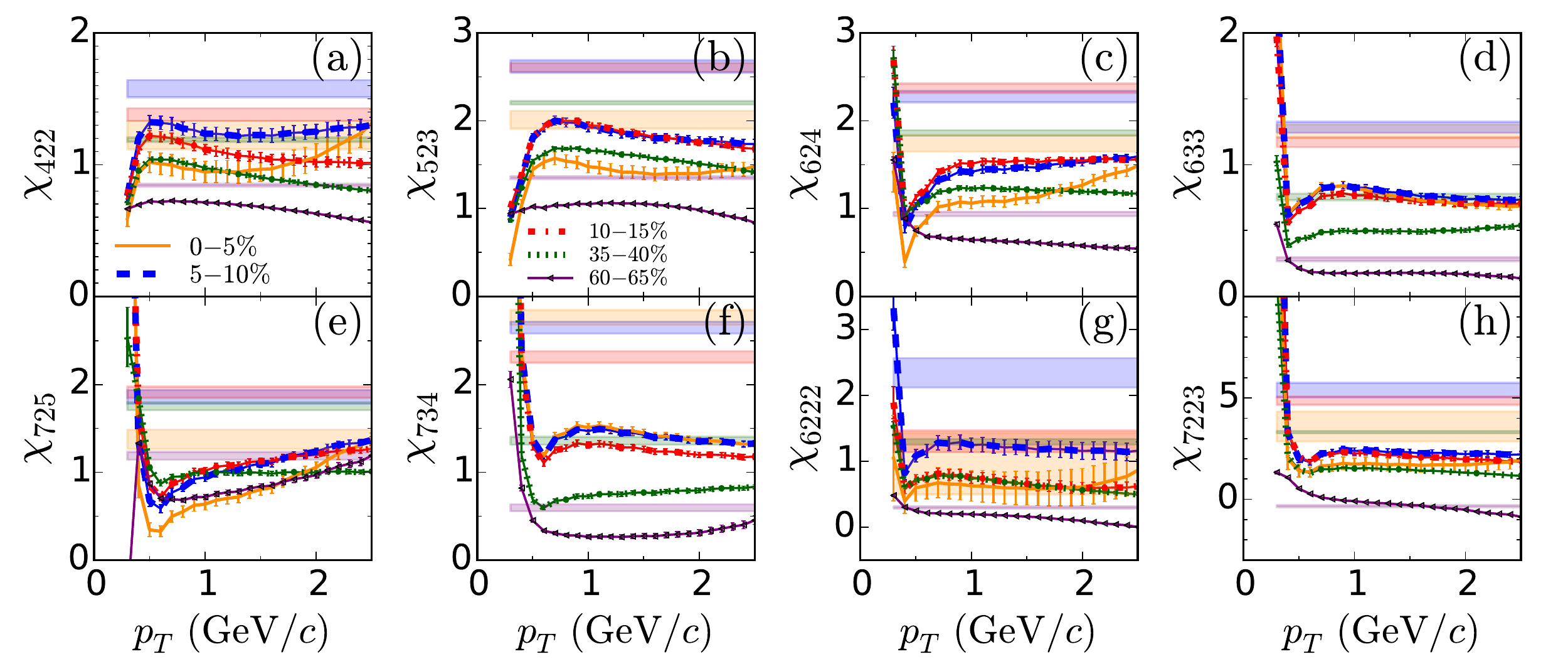}
    \caption{(Color online)
    \baselineskip=14pt
   $p_T$ dependence of the mode coupling coefficients for $p_T$-differential flows from viscous 
   hydrodynamics with $\eta/s=0.08$. Lines of different colors and styles represent different collision
   centralities. Each 5\% centrality bin contains 3000 events. For comparison, shaded bands of the 
   same color as the lines show the mode coupling coefficients for the integrated flows corresponding 
   to the same centrality.
     \label{F-IV1}
}
\end{figure*}
%

As discussed in the preceding section, the $p_T$-differential flows exhibit qualitatively similar mode coupling effects as the integrated flows. To quantify them we compute the mode coupling coefficients for the differential flows according to Eqs.~(\ref{chi}) and show their $p_T$ dependence in Fig.~\ref{F-IV1}. The $p_T$-differential mode coupling coefficients show similar centrality dependence as the integrated ones (shown as colored shaded bands) and generally have only a weak dependence on $p_T$, except at small $p_T$. The strong variation of the mode coupling coefficients at small $p_T$ is related to the similarly strong $p_T$-dependence of the event-plane correlators shown in Fig.~\ref{F-II1} and can again be attributed to the large variance of the flow angles $\Psi_n$ at small $p_T$. 

In the $p_T$ regions where $p_T$-differential mode coupling coefficients show only weak $p_T$ dependence their magnitudes are generically smaller than those of the integrated flows, for almost all modes and and for all collision  centralities studied here. To understand this intuitively let us consider the case of quadrangular flow in the approximation where the non-linear mode coupling contribution dominates:
\begin{equation}
\begin{split}
  V_4&\approx\chi_{422} V_2^2,  \quad \quad \quad V_4(p_T)\approx\chi_{422}(p_T) V^2_2(p_T),\\
  \chi_{422}&\approx\frac{V_4}{V_2^2} =\frac{v_4 e^{i4\Psi_4}}{ v_2^2 e^{i4\Psi_2}}=\frac{v_4}{v_2^2}\\
  &=\frac{\{v_4(p_T)\}}{\{v_2(p_T)\}^2}=\frac{\{\chi_{422}(p_T)\,v^2_2(p_T)\}}{\{v_2(p_T)\}^2}
  \end{split}
\label{derive1}
\end{equation}
Here $\{...\}$ denotes symbolically the averaging of the differential flow over $p_T$. If $\chi_{422}(p_T)$ is independent of $p_T$ (which according to Fig.~\ref{F-IV1} is approximately true for $p_T \geq0.5$\,GeV/$c$) then
\begin{equation}
\chi_{422}\approx\chi_{422}(p_T) \frac{ \{ v^2_2(p_T) \}}{\{v_2(p_T)\}^2} \geq \chi_{422}(p_T).
\label{derive2}
\end{equation}

As observed above when discussing the $v_m-v_n$ correlations, the $k_n$ values associated with the integrated flows are also larger than those of the $p_T$-differential flows. In fact, both the $\chi$ and $k$ coefficients describe mode coupling effects, and hence they are tightly connected. Taking $n=4$ as an example, this is illustrated by combining Eqs.~(\ref{vmvn}) and (\ref{VnDecomposition}) as follows:
\begin{eqnarray}
\label{derive3}
       \langle v_4^2 \rangle &=& (v_4^0)^2 + k_4^2 \langle v_2^2 \rangle^2 
\nonumber\\ 
        &=& \langle v_{4L}^2 \rangle + \chi_{422}^2 \langle v_2^4 \rangle
         =  \langle v_{4L}^2 \rangle + \chi_{422}^2 \Bigl(\langle v_2^2\rangle^2{+}\sigma^2_{v_2^2}\Bigr).
\end{eqnarray}
Here $\sigma^2_{v_2^2}$ is the variance of $v_2^2$, and we used $\langle V_{4L} V^{*2}_2\rangle \approx 0$ \cite{Yan:2015jma,Qian:2016fpi}.  Using the first of these equations, a similar argument as in Eqs.~(\ref{derive1},\ref{derive2}) provides support for our observation that $k_4\geq k_4(p_T)$.

Using the decomposition (\ref{VnDecomposition}) for the higher-order flows in the form $\langle v_n^2 \rangle = v_{nL}^2 + v_{nM}^2$ we can separate the linear and mode coupling terms as follows:
\begin{eqnarray}
v_{4L} &=& \sqrt{\langle v_4^2 \rangle - \frac{|\mathrm{Re} \langle V_4 V_2^{*2} \rangle|^2}{\langle v_2^4 \rangle}},\quad 
v_{4M} = \frac{|\mathrm{Re} \langle V_4 V_2^{*2} \rangle|}{\sqrt{\langle v_2^4 \rangle}},
\nonumber\\
v_{5L} &=& \sqrt{\langle v_5^2 \rangle - \frac{|\mathrm{Re} \langle V_5 V_2^* V_3^* \rangle|^2}{\langle v_2^2 v_3^2 \rangle}},\quad 
v_{5M} = \frac{|\mathrm{Re} \langle V_5 V_2^* V_3^* \rangle|}{\sqrt{\langle v_2^2 v_3^2 \rangle}}.
\label{v4v5}
\end{eqnarray}
%

%
\begin{figure}[!hbt]
    \includegraphics[width=\linewidth]{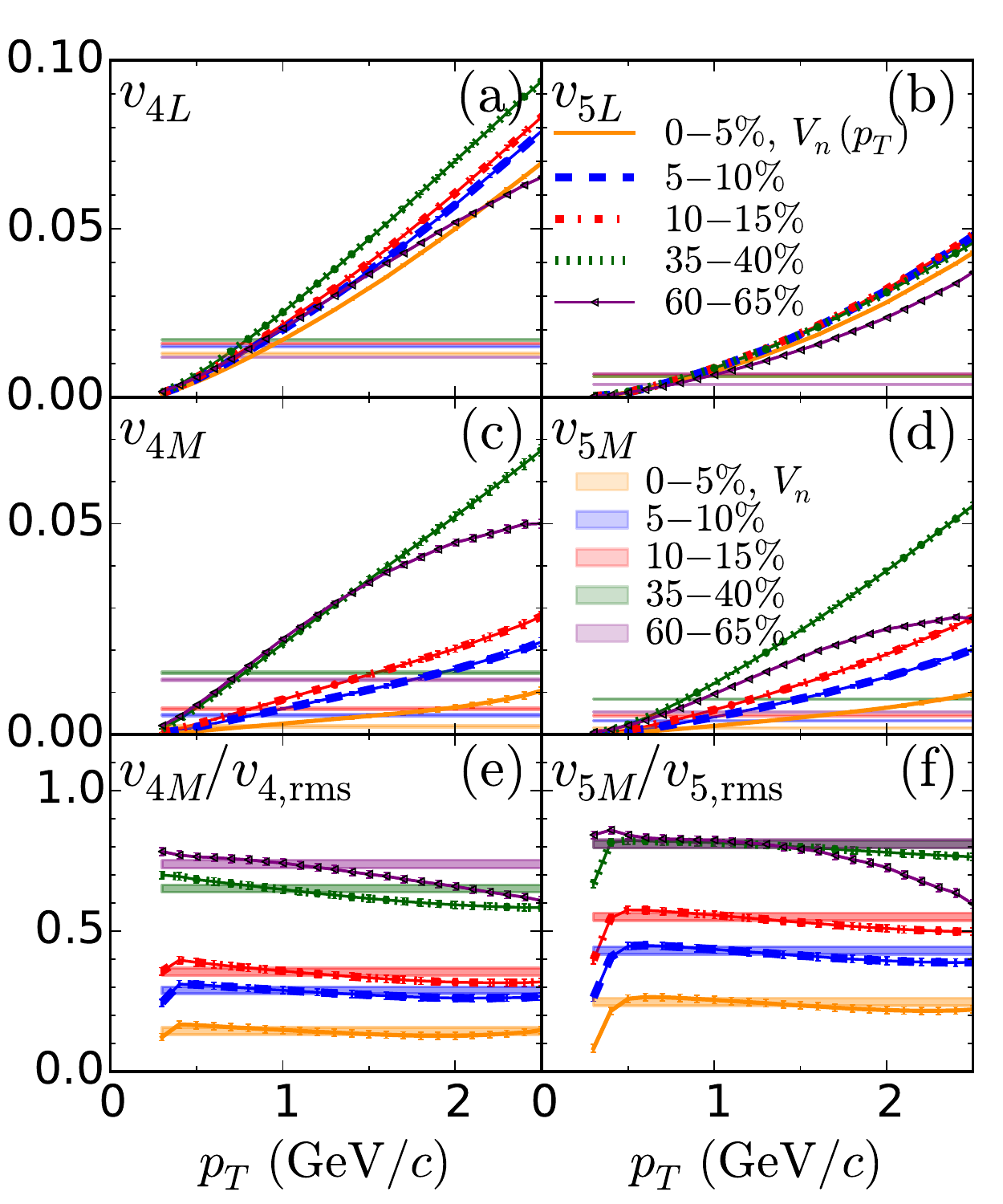}
    \caption{(Color online)
    \baselineskip=14pt
    $p_T$ dependence of $v_{nL}$, $v_{nM}$ and $v_{nM}/v_{n,\mathrm{rms}}$ for $n=4, 5$ from viscous 
    hydrodynamics with $\eta/s=0.08$. Lines with different colors and symbols represent different 
    centrality classes, with 3000 events in each 5\% centrality bin. For comparison, the colored 
    shaded bands show the corresponding values for the $p_T$-integrated flows.
     \label{F-IV2}
}
\end{figure}
%

Figure~\ref{F-IV2} shows the linear and mode-coupling contributions to the $p_T$-differential flows, $v_{nL}$ and $v_{nM}$ as defined in Eqs.~(\ref{v4v5}), together with the ratio of the latter with $v_{n,\mathrm{rms}}(p_T)\equiv\sqrt{\langle v_n^2(p_T)\rangle}$ (which indicates the relative importance of the mode-coupling terms to the $p_T$-differential flows), as functions of $p_T$ for different collision centralities. Similar to what was observed earlier for the $p_T$-integrated flows \cite{Qian:2016fpi}, the linear and mode-coupling contributions to the differential flows exhibit opposite centrality dependences:
the linear terms depend relatively weakly on centrality whereas the mode-coupling terms increase rapidly with increasing impact parameter. 

Comparing the definition (\ref{EP_SP}) of the event plane correlations with Eqs.~(\ref{v4v5}) for the mode-coupling contributions $v_{nM}$ one sees that for $n=4, 5$ the correlation of the $n$th order event plane with those of lower harmonic order is, in fact, given by the fraction $v_{nM}/v_{n,\mathrm{rms}}$ of the $n$th order rms flow $v_{n,\mathrm{rms}}$ contributed by mode-coupling effects, plotted in the bottom row of Fig.~\ref{F-IV2}.\footnote{%
	\baselineskip 12pt
	More precisely, the two observables are equal up to a sign. For example, 
	$\langle \cos(4(\Psi_2{-}\Psi_4)) \rangle\{\mathrm{SP}\}=\mathrm{sign}(\chi_{422}) \,
	   v_{4M}/v_{4,\mathrm{rms}}$.
	Since the two event plane correlators discussed in this paper, 
	$\langle \mathrm{cos}(4(\Psi_2-\Psi_4)) \rangle\{\mathrm{SP}\}$ and 
	$\langle \mathrm{cos}(2\Psi_2 + 3\Psi_3 - 5\Psi_5)\rangle\{\mathrm{SP}\}$, are mostly positive 
	(except at small $p_T$ where the event planes fluctuate strongly) we omit the sign for short.}
\begin{equation}
\begin{split}
   \langle \mathrm{cos}(4(\Psi_2{-}\Psi_4)) \rangle \{\mathrm{SP}\}
   &= \frac{\langle V_4 V_2^{*2} \rangle}{\sqrt{\langle v_4^2 \rangle \langle v_2^4 \rangle}}
   =\frac{v_{4M}}{v_{4,\mathrm{rms}}},
\\
    \langle \mathrm{cos}(2\Psi_2{+}3\Psi_3{-}5\Psi_5)) \rangle\{\mathrm{SP}\} 
    &= \frac{\langle V_5 V_2^* V_3^* \rangle}{\sqrt{\langle v_5^2 \rangle \langle v_2^2 \rangle \langle v_3^2 \rangle}}
    = \frac{v_{5M}}{v_{5,\mathrm{rms}}}.
\end{split}
\label{EP_ratio}
\end{equation}
That means that for $n=4, 5$ the mode coupling contributions to the flow magnitudes are caused by correlations between the $n$th-order and lower-order event planes. By implication, the smallness of the event plane correlations shown in Fig.~\ref{F-II1} near $p_T=0$ should lead to similarly small mode coupling contributions to $v_{4,5}(p_T)$ at small $p_T$. While this $p_T$ region is not shown in Fig.~\ref{F-IV2}, for reasons explained in Sec.~\ref{sec2}, the bottom panels in Fig.~\ref{F-IV2} indicate a steep drop of the mode coupling contributions to $v_4$ and $v_5$ below $p_T\sim0.5$\,GeV/$c$. At higher $p_T$, the approximate $p_T$-independence of the event plane correlations shown in Fig.~\ref{F-II1} is reflected in the flatness of $v_{nM}(p_T)/v_{n,\mathrm{rms}}(p_T)$ shown in the bottom panels of Fig.~\ref{F-IV2}.

Since, except for small $p_T$, the differential flows $v_{4,5}(p_T)$ receive a mode-coupling contribution that is almost independent of $p_T$, the mode-coupling contribution to their $p_T$-integrated analogues is very similar. The shaded bands in Figs.~\ref{F-IV2}e,f show this. The above connection between mode-coupling effects and event plane correlations thus provides an explanation for the similarity of the strengths of the event plane correlations for $p_T$-integrated and $p_T$-differential flows noted in the discussion of Fig.~\ref{F-II1} in Sec.~\ref{sec2}.

\section{Summary and conclusions}
\label{sec5}

Using viscous hydrodynamics as a model for the dynamical evolution of Pb-Pb collisions at the LHC we presented a first systematic study of the correlations between different harmonic orders of the $p_T$-differential anisotropic flows of charged hadrons. We identified nonlinear mode coupling contributions to the differential flow, studied their $p_T$ and centrality dependence and compared them with those for the $p_T$-integrated flows. We identified correlations with lower-order event planes as the main contributor to the mode coupling effects seen in the magnitudes of higher-order harmonic flows. Except for very low $p_T$, the mode coupling fraction depends very weakly on transverse momentum, and this is reflected in event plane correlations between the differential flows that are close to those of the integrated flows and largely independent of $p_T$. At very low $p_T$ they exhibit strong $p_T$ dependence, caused by large, $p_T$-dependent fluctuations of the flow angle. These event plane fluctuations destroy the mode coupling contributions to the higher-order flow magnitudes at low $p_T$, by averaging them away. Correlations between the magnitudes of the $p_T$-differential flows of different order have similar strength and centrality dependence as those between the corresponding integrated flows. The mode-coupling coefficients extracted from a two-component fit using event-shape engineering techniques were found to be smaller for the $p_T$-differential flows than those for the integrated flows. This observation has a simple explanation as described in Sec.~\ref{sec4}.  The linear part of the two-component fit was shown to reflect the linear hydrodynamic flow response to the cumulant-based initial eccentricities but not to their standard moment-based analogues.

\begin{acknowledgments}
The authors thank Jia Liu, Christopher Plumberg and Jianyi Chen for discussions. The research of UH was supported by the U.S. Department of Energy, Office of Science, Office for Nuclear Physics under Award \rm{DE-SC0004286}. Computing resources were generously provided by the Ohio Supercomputer Center \cite{OhioSupercomputerCenter1987}. 
\end{acknowledgments}

\appendix
\vspace*{-3mm}

\section{Discussion of $V_{nL}$}
\label{appa}

Usually, the eccentricities of the initial spatial distributions of energy or entropy are defined as moments with weight $r^n$: $\mathcal{E}_n=\epsilon_n e^{in\Phi_n} \equiv - \frac{\langle r^n e^{in \phi} \rangle}{\langle r^n \rangle}$ (for $n > 1)$. The authors of \cite{Teaney:2012ke, Teaney:2013dta} suggested a different set $\mathcal{E}'_n$ of eccentricity coefficients using spatial cumulants:
\begin{equation}
\begin{split}
 \mathcal{E}_2' &\equiv \epsilon_2 e^{i2\Phi_2} = \mathcal{E}_2, \quad \quad \quad 
  \mathcal{E}_3' \equiv \epsilon_3 e^{i3\Phi_3} = \mathcal{E}_3, \\
 \mathcal{E}_4' &\equiv \epsilon_4' e^{i4\Phi_4'} \equiv - \frac{\langle z^4 \rangle - 3 \langle z^2 \rangle ^2}{\langle r^4 \rangle}  = \mathcal{E}_4 + \frac{3\langle r^2 \rangle ^2}{\langle r^4 \rangle} \mathcal{E}_2^2, \\
 \mathcal{E}_5' &\equiv \epsilon_5' e^{i5\Phi_5'} \equiv - \frac{\langle z^5 \rangle - 10 \langle z^2 \rangle \langle z^3 \rangle}{\langle r^5 \rangle}  = \mathcal{E}_5 + \frac{10\langle r^2 \rangle \langle r^3 \rangle}{\langle r^5 \rangle} \mathcal{E}_2 \mathcal{E}_3, \\
 \mathcal{E}_6' &\equiv \epsilon_6' e^{i6\Phi_6'} \equiv - \frac{\langle z^6 \rangle - 15 \langle z^2 \rangle \langle z^4 \rangle - 10 \langle z^3 \rangle ^2 + 30 \langle z^2 \rangle ^3}{\langle r^6 \rangle}  \\
          &= \mathcal{E}_6 + \frac{15\langle r^2 \rangle \langle r^4 \rangle}{\langle r^6 \rangle} \mathcal{E}_2 \mathcal{E}_4 + \frac{10\langle r^3 \rangle ^2}{\langle r^6 \rangle} \mathcal{E}_3^2 + \frac{30\langle r^2 \rangle ^3}{\langle r^6 \rangle} \mathcal{E}_2^3.
\end{split}
\label{EcnNew}
\end{equation}
Here $z \equiv x+iy = re^{i\phi}$. Note that $\mathcal{E}_1$ (which is not used in our discussion) has a different definition.

\begin{figure}[!hbt]
    \includegraphics[width=\linewidth]{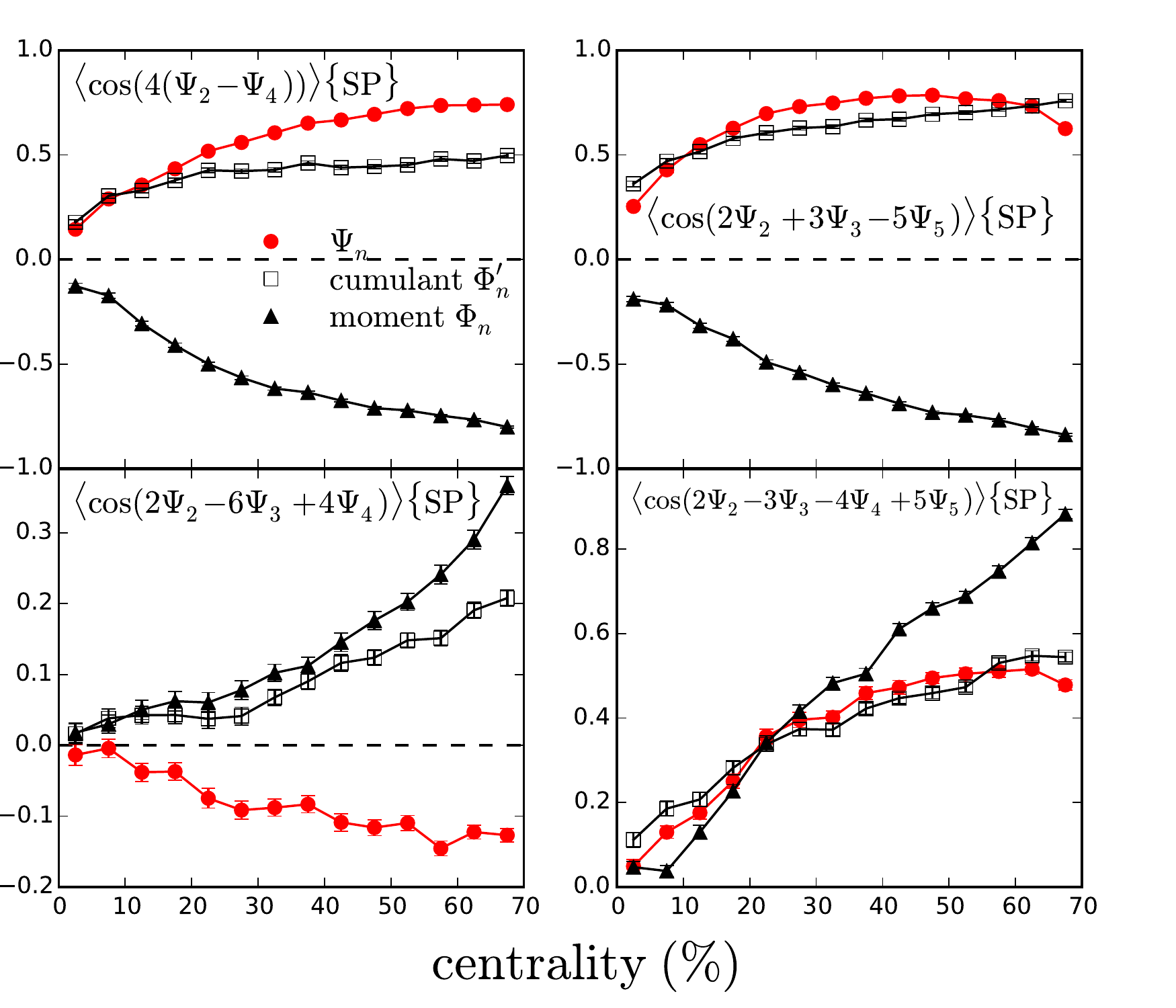}
    \caption{(Color online)
    \baselineskip=14pt
    Event plane correlations and the corresponding participant plane correlations from viscous hydrodynamics 
    with MC-Glauber initial conditions and $\eta/s=0.08$. 42000 minimum bias (0-70\% centrality) events are 
    cut into 14 equal centrality classes according to their multiplictiy. Red circles represent event plane 
    correlations between flow vectors integrated over the $p_T$ range $p_T\in [0.3, 3.5)$\,GeV/$c$, black 
    triangles represent the corresponding participant plane correlators with moment-defined eccentricities 
    while black hollow squares represent the corresponding participant plane correlators of the cumulant-defined 
    eccentricity vectors. 
     \label{Fa2}
}
\end{figure}

By defining eccentricities using cumulants instead of moments one subtracts contributions from lower order $z$ correlations. This led Teaney and Yan to suggest \cite{Teaney:2012ke, Teaney:2013dta} that the linear hydrodynamic response contribution to higher order flows should be linearly proportional to the cumulant-defined eccentricities and not to the traditional moment-defined ones. They also used this hypothesis to successfully explain the experimentally observed event plane correlators in terms of linear response to the corresponding participant plane correlators, except for one event plane correlator: $\langle \mathrm{cos}(2\Psi_2 - 6\Psi_3 +4\Psi_4 ) \rangle$ \cite{Teaney:2012gu,Teaney:2013qka}. Figure~\ref{Fa2} shows that we agree with their findings. Like them, we do not have an explanation for the apparent non-linearity of the response leading to the 2-3-4 flow correlator.

%

\begin{figure}[!hbt]
    \includegraphics[width=\linewidth]{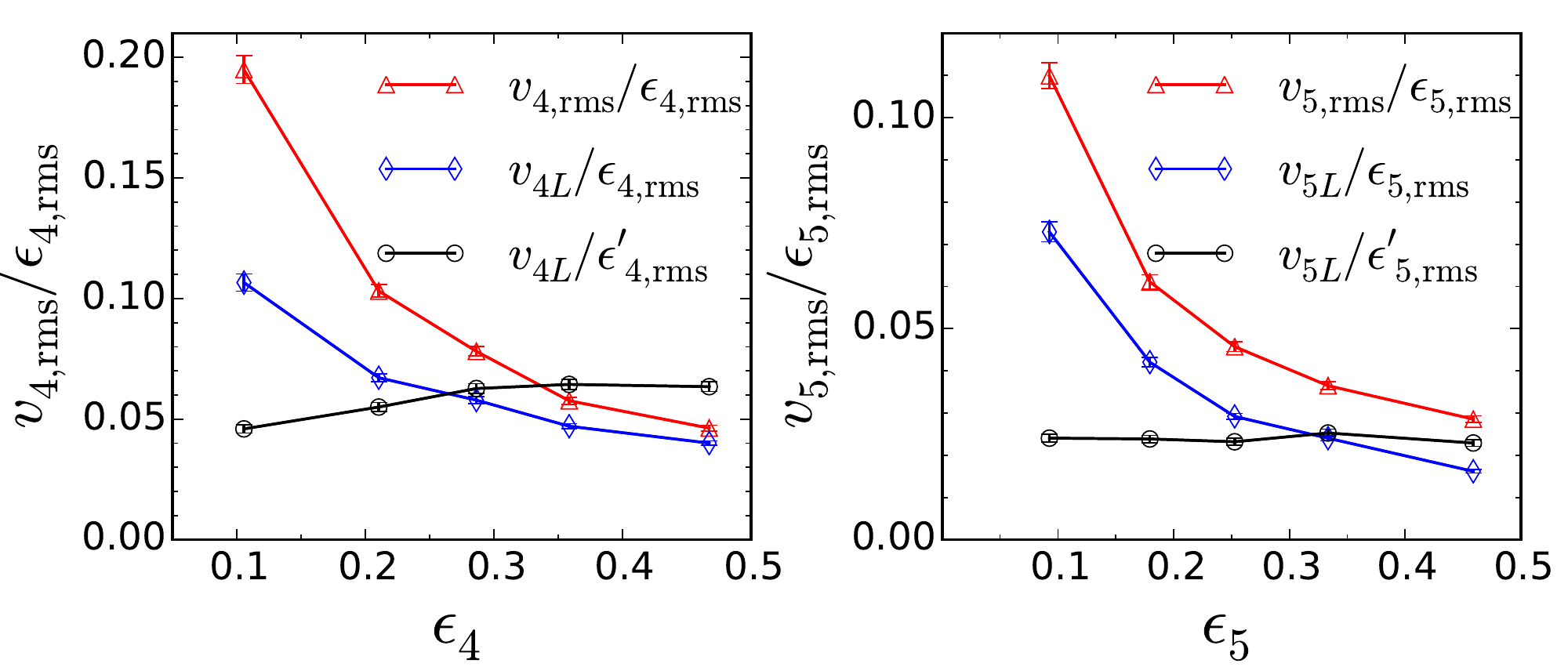}
    \caption{(Color online)
    \baselineskip=14pt
    Check of the linearity between $v_n$ and $\epsilon_n$, using 2000 ideal hydrodynamic events at 
    40-50\% centrality. Red triangles represent $v_n\{2\}/\epsilon_n\{2\}$, blue diamonds represent 
    $v_{nL}/\epsilon_n\{2\}$ and black circles are $v_{nL}/\epsilon_n'\{2\}$. The left panel is for $n=4$ while 
    the right panel is for $n=5$.
     \label{Fa1}
}
\end{figure}
%

To further examine the (non-)linearity of the hydrodynamic response, we use the ratio between flow and eccentriciy as a function of eccentricity. Since in Eqs.~(\ref{v4v5}) the square of the magnitude of the linear response term is averaged over events, we use the root mean square eccentricities for normalization:
\begin{equation}
\begin{split}
 v_{n,\mathrm{rms}}=\sqrt{\langle v_n^2 \rangle},\\
 \epsilon_{n,\mathrm{rms}} = \sqrt{\langle |\mathcal{E}_n|^2 \rangle},\\
 \epsilon'_{n,\mathrm{rms}} = \sqrt{\langle |\mathcal{E}'_n|^2 \rangle},
\end{split}
\label{Illustrate}
\end{equation}
with $\mathcal{E}'_n$ from Eqs.~(\ref{EcnNew}). In Fig.~\ref{Fa1} we plot the ratios $v_n/\epsilon_n$ as functions of $\epsilon_n$ for different definitions of numerator and denominator as described in the legend, for $n=4$ and 5, using 2000 ideal hydrodynamic events at 40-50\% centrality. We find that, different from $v_n\{2\}/\epsilon_n\{2\}$ (shown as red triangles) and $v_{nL}/\epsilon_n\{2\}$ (shown as blue diamonds), $v_{nL}/\epsilon'_n\{2\}$ (shown as black circles) is almost independent of the $\epsilon_n$ used in the denominator, suggesting that $V_{nL}$ is indeed the linear response to $\mathcal{E}'_n$.


\end{document}